\documentclass[floatfix,prl,twocolumn,superscriptaddress,sn-aps,longbibliography]{revtex4-1}
\usepackage[utf8]{inputenc}

\usepackage{graphicx,epsfig,amsfonts}
\usepackage{bm}
\usepackage{amsmath}
\usepackage{amssymb}
\usepackage{float}
\usepackage{xcolor}
\usepackage[colorlinks=true,citecolor=blue]{hyperref} 
\usepackage{tabularx}

\DeclareGraphicsExtensions{.png,.jpg,.eps}
\usepackage{xcolor}

\newcommand{\beq}{\begin{eqnarray}}
\newcommand{\eeq}{\end{eqnarray}}
\renewcommand{\H} {\mathcal{H}}

\begin{document}

\title{Hamiltonian learning with real-space impurity tomography in topological moir\'e superconductors
}

\author{Maryam Khosravian} 
\thanks{These two authors contributed equally}
\affiliation{Department of Applied Physics, Aalto University, 02150 Espoo, Finland}

\author{Rouven Koch}
\thanks{These two authors contributed equally}
\affiliation{Department of Applied Physics, Aalto University, 02150 Espoo, Finland}

\author{Jose L. Lado}
\affiliation{Department of Applied Physics, Aalto University, 02150 Espoo, Finland}


\begin{abstract}
Extracting Hamiltonian parameters from available experimental data is a challenge in quantum materials.
In particular, real-space spectroscopy methods such as scanning tunneling spectroscopy allow probing electronic states
with atomic resolution, yet even in those instances extracting the effective Hamiltonian is an open challenge.
Here we show that impurity states in modulated systems provide a promising approach to extracting non-trivial Hamiltonian parameters of a quantum material. We show that by combining the real-space spectroscopy of different impurity locations in a moir\'e topological superconductor, modulations of exchange and superconducting parameters can be inferred via machine learning.
We demonstrate our strategy with a physically-inspired harmonic expansion combined with a fully-connected neural network that we benchmark against a conventional convolutional architecture.
We show that while both approaches allow extracting exchange modulations, only the former approach allows
inferring the features of the superconducting order. Our results demonstrate the potential of machine learning methods to extract Hamiltonian parameters by real-space impurity spectroscopy as local probes of a topological state.
\end{abstract}

\maketitle

\section{Introduction}

Learning Hamiltonian parameters from experimental data is one of the most critical open problems in order
to bring together experiments with theoretical models\cite{PhysRevX.8.021026,PhysRevX.8.031029,PhysRevLett.122.150606,PhysRevResearch.4.033223,PhysRevResearch.1.033092,2022arXiv221207893K,PhysRevResearch.3.023246,Yang2021}. Conventionally, phenomenological models to account for experimental data are developed on a case by case basis. In certain instances, obtaining the Hamiltonian parameters of the model can be done by fitting specific features of the data\cite{PhysRevLett.111.127203,Yan2015}. 
Many-body Hamiltonians with local interactions 
can be extracted local observables by exploiting
time evolution and quantum Hamiltonian tomography, a strategy
demonstrated theoretically and experimentally\cite{Li2020,PhysRevLett.122.020504,PhysRevLett.130.200403,PhysRevA.105.023302,Kokail2021,Gentile2021}. 
However, in some instances, no simple fitting procedure nor time-dependent measurements
can be performed to extract Hamiltonian parameters. 
Machine learning methods have risen as a powerful approach to extract subtle features of data, and in particular they have become successful in tackling inverse problems in quantum materials\cite{PhysRevLett.102.187203,Carrasquilla2020,PhysRevA.80.022333,PhysRevA.98.032114,Hincks2018,Wang2017,Gebhart2023,2023arXiv230410852K,10.21468/SciPostPhysCore.6.2.030,2023arXiv231107253A,PRXQuantum.2.020303,Sobral2023}.

\begin{figure}
    \centering
    \includegraphics[width=.97\linewidth]{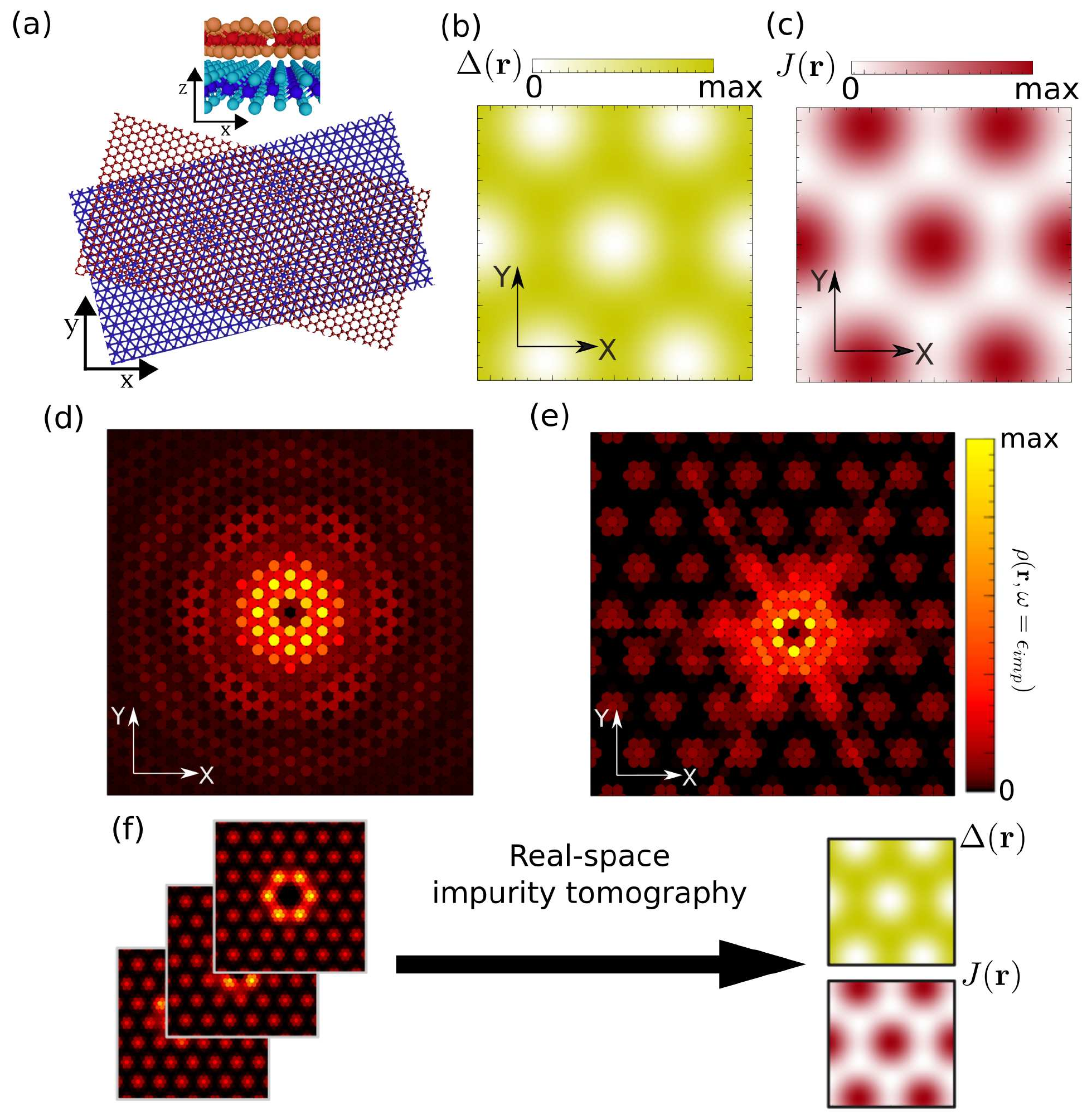}
    \caption{Schematic of the moir\'e system (a),
    where panel (b) denotes the modulation of the superconducting and (c) the modulation of the exchange parameters.
    In the absence of moir\'e, impurities induced in-gap states in the topological superconducting state
    (d), which upon the existence of the moir\'e pattern gives rise to an interference effect (e).
    Using the spectroscopy of impurities in different locations, a machine learning algorithm extracts the
    exchange and superconducting profiles (f).
    }
    \label{fig:schematic}
\end{figure}

Van der Waals heterostructures are a paradigmatic system in which, thanks to their tunability, a variety of
quantum Hamiltonians can be emulated\cite{Andrei2021}. 
Specifically, the capability of combining two-dimensional materials with
radically different properties allows engineering systems simultaneously
hosting antagonist order\cite{Gibertini2019,Liu2016}, with the paradigmatic example of
van der Waals heterostructures combining 2D magnets
and superconductors\cite{Kezilebieke2020,Kezilebieke2021,Kezilebieke2022,Persky2022,Ai2021}. 
Beyond the different coexisting orders, van der Waals materials
bring a unique feature to heterostructures, the emergence of a moir\'e pattern\cite{PhysRevLett.99.256802,PhysRevB.82.121407,Bistritzer2011,PhysRevB.93.235153,PhysRevB.92.075402}. The moir\'e pattern stems from the structural modulation in real space due to a twist angle between lattices. Most importantly, the moir\'e modulation leads to a spatial modulation of all the Hamiltonian parameters. This modulation is the driving force behind a variety of phenomena, including topological and correlated states in van der
Waals heterostructures\cite{Cao20181,Cao2018, Yankowitz2019, Lu2019, Chen2019,Arora2020,tripletCao,Sharpe2019,Lu2019, Xie2021,Wong2020,Zondiner2020, Saito2021, Datta2023}.
However, extracting the values of these modulations
in moir\'e system is an open problem, as their effect on the electronic structure is greatly challenging to disentangle. Interestingly, the coexistence of a moir\'e pattern and local impurities opens up a new strategy to
infer electronic parameters due to their non-trivial interplay\cite{khosravian2022impurity,2022arXiv221101038S,PhysRevMaterials.3.084003,PhysRevB.99.245118,2023arXiv230403018B}.

In our manuscript, we put forward a method to extract Hamiltonian parameters of unconventional moir\'e superconductors
using machine learning and impurity engineering. Specifically, we show that thanks to the moir\'e pattern,
local impurities give rise to dramatically different electronic excitations depending on the location of the moir\'e pattern. This dependency allows us to directly infer moir\'e modulations via a machine learning algorithm that takes as input excitations for different impurity locations. We demonstrate that our algorithm allows extracting modulation strengths of exchange proximity and superconducting order, two parameters that cannot be
directly extracted from a measured local density of states. 
We further address the robustness of our algorithm to noise, showing that Hamiltonian parameters can be extracted in an experimentally realistic scenario.

\section{Methods}

\subsection{Model}

We consider an artificial two-dimensional superconductor
obtained by combining a two-dimensional ferromagnet and a two-dimensional
superconductor\cite{Kezilebieke2020,Kezilebieke2021,Kezilebieke2022}
as shown in Fig.~\ref{fig:schematic}a.
The electronic structure of the heterostructure
is modeled with an atomistic Wannier orbital site forming a triangular lattice,
where the moir\'e pattern is incorporated in the
modulation of the Hamiltonian
parameters\cite{PhysRevB.104.195156,PhysRevB.104.075126,PhysRevLett.124.136403,PhysRevB.103.085109,Karpiak2019}.
The full Hamiltonian takes the form

\begin{equation}
\H_{0} = 
\H_{\text{kin}}
+
\H_J +
\H_{R}  +
\H_{\text{SC}}
\label{eq:h1}
\end{equation}
with
$
\H_{\text{kin}} = 
t \sum_{\langle ij\rangle,s} c^{\dagger}_{i, s}c_{j,s} +\mu \sum_{i}c^{\dagger}_{i, s}c_{i,s}
$
with $c_{n,s}^{\dagger}$( $c_{n,s}$) the creation (annihilation) fermionic operator 
with spin $s$ in site $n$. 
The combination of exchange coupling, Rashba spin-orbit coupling and
superconducting proximity gives rise to topological superconducting
state when combined on the right footing\cite{Beenakker2013,Alicea2012,Li2016,PhysRevLett.114.236803,Mnard2017,Pyhnen2018,Kezilebieke2020}.
We focus on the regime giving rise to topological superconducting states $C=1,2,3$,
which arise when taking the chemical potential crossing the $\Gamma, K, M$ points,
respectively.
The hopping is controlled by $t$, the chemical potential by $\mu$
and $\langle i, j\rangle $ runs
over nearest neighbors.
The term
$
\H_{R} = 
    i\lambda_R  \sum_{\langle ij \rangle, s s^{'}} 
    \mathbf{d}_{ij} 
    \cdot 
    \mathbf{\sigma} ^{s,s^{'}} c^{\dagger}_{i, s}c_{j,s^{'}}
$
is the Rashba spin-orbit coupling
arising from mirror symmetry breaking
at the interface\footnote{We take $\lambda_R = 0.2t$ for all calculations.}, with
$\mathbf \sigma$  the spin Pauli
matrices, $\lambda_R$ controls the spin-orbit coupling constant
and
$\mathbf{d}_{ij} = (\mathbf{r}_i - \mathbf{r}_j)\times \hat z$.

We now focus on the main terms that we will extract with our procedure, the modulated exchange and modulated
superconductivity. The exchange coupling is included in a term

\begin{equation}
    \H_J = 
    \sum_{i,s,s'} J(\mathbf r) \sigma_z^{s,s'} c^{\dagger}_{i,s}c_{i,s'} 
    \label{ej}
\end{equation}

\begin{figure}
    \centering
    \includegraphics[width=\linewidth]{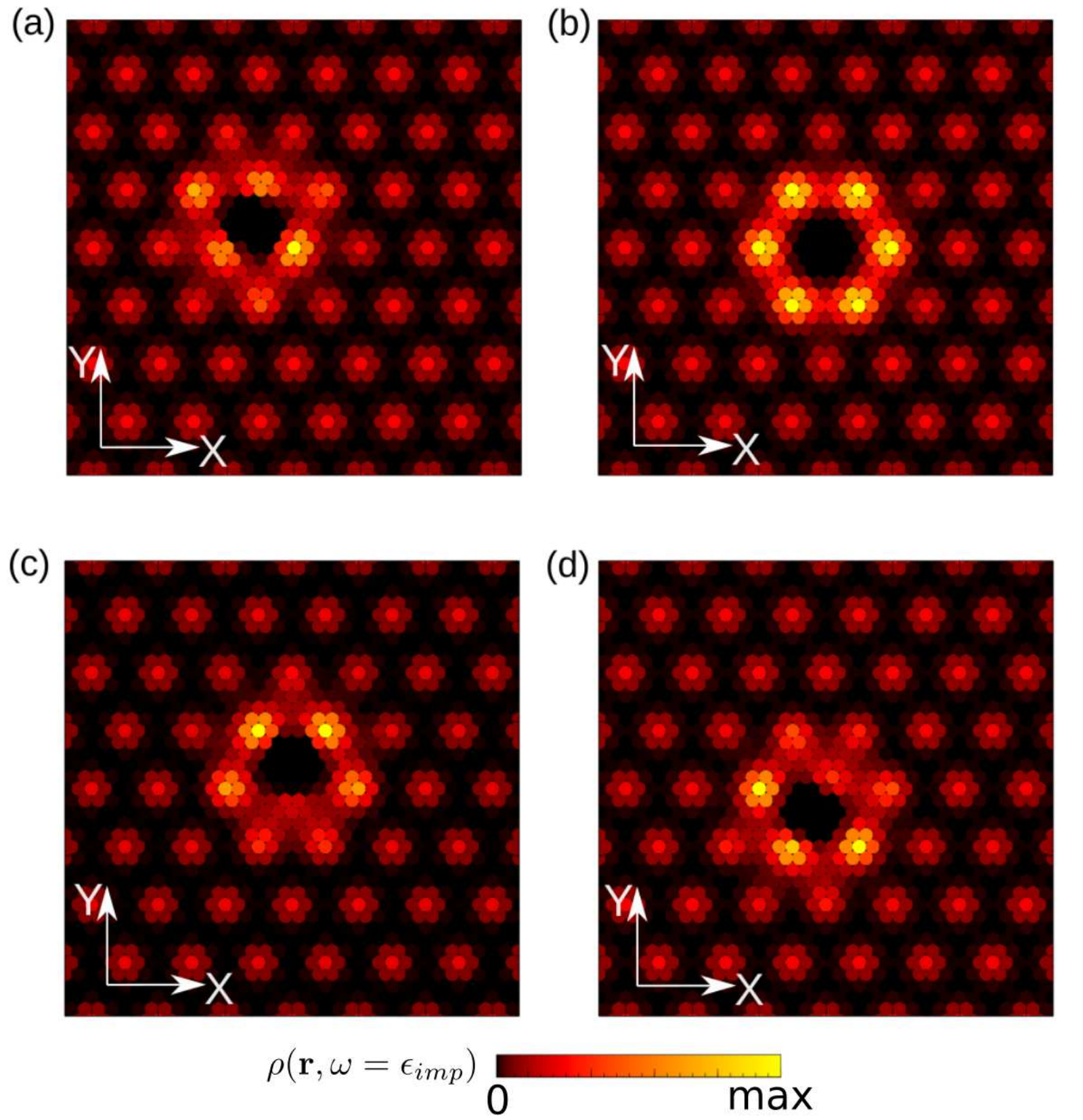}
    \caption{Local density of states of the in-gap state
    for a single impurity in a moir\'e
    topological superconductor for different impurity locations (a-d).
    It is observed that depending on the location of the impurity in the moir\'e,
    the in-gap states show different interference
    patterns with the moir\'e potential.
    These different interference patterns
    reflect the underlying modulations of the Hamiltonian,
    and allow us to extract Hamiltonian modulation from real-space spectroscopy.
    In the absence of the moir\'e pattern, all the impurity locations would
    show the same profile of the in-gap state.}
    \label{fig:ldos}
\end{figure}

and the superconducting order by

\begin{equation}
\H_{\text{SC}} = 
    \sum_{i} \Delta (\mathbf r) c^{\dagger}_{i,\uparrow}c^{\dagger}_{i,\downarrow}+h.c.
   \label{esc}
\end{equation}
where $J(\mathbf{r})$ and $\Delta(\mathbf{r})$ parameterize the exchange coupling and 
induced s-wave superconductivity. 
The competition between the exchange field and the superconducting
order leads to an induced modulated
superconductivity stemming from the originally
modulated exchange\cite{DeGennes2018,PhysRevB.107.174522,RevModPhys.77.935}.
We take a modulation parametrized as
$
f(\mathbf r) = c_0 + c_1 \sum _{n=1}^{3}\cos({\it {R^{n}}\mathbf{q} \cdot \mathbf r}) 
\label{eq:modul}
$ where ${\mathbf q}$ is the moir\'e superlattice wave vector, 
and $R_{n}$ is the rotation matrix
conserving $C_3$ symmetry.
The parameters $c_0, c_1$ are defined so that $\langle f (\mathbf r) \rangle = 0$. 
The spatial 
profiles $J(\mathbf{r})$ and $\Delta(\mathbf{r})$ 
are written in terms
of the previous spatial dependence as

\begin{align}
J(\mathbf r) = J_{0}+ \delta_{J} f(\mathbf r) \nonumber \\
\Delta(\mathbf r) = \Delta_{0}+ \delta_{\Delta} (1-f(\mathbf r))
\label{eq:amplitude}
\end{align}

$J_{0}$ and $\Delta_{0}$ parametrize the average magnitude of the modulated exchange and superconducting profiles,
whereas $\delta_{J}$ and $\delta_{\Delta}$ control
the amplitude of the moir\'e modulation.
For the sake of concreteness we take $J_0=\lambda_R$ and $\Delta_0=\lambda_R/2$. 
The relative signs of $f(\mathbf r)$ in $J(\mathbf r)$ and $\Delta(\mathbf r)$ are taken so that
when the superconducting order is maximum, the exchange is minimum (Fig.~\ref{fig:schematic}bc).

Local non-magnetic impurities are included
adding a potential scattering term
of the form

\begin{equation}
    \H_{\text{imp}} = 
    W \sum_s c^{\dagger}_{n,s}c_{n,s}
\end{equation}

where $\H_{\text{imp}}$ defines the impurity Hamiltonian at site $n$ with an on-site 
potential $w$. We focus on the strong impurity limit in which
the site is effectively removed from the low energy manifold
taking $W=100t$. 
The full Hamiltonian of the defective system takes the form

\begin{equation}
    \H = \H_0 + \H_{\text{imp}}
\end{equation}

The impurity gives rise to an in-gap state both in the absence (Fig.~\ref{fig:schematic}d)
and presence (Fig.~\ref{fig:schematic}(e)) of a moir\'e modulation.
It is important to note that, due to the presence of the moir\'e pattern,
the impact of an impurity depends on its location with respect to the moir\'e modulation.
Specifically, as shown in the local density of states of
Fig.~\ref{fig:ldos}, the interference between the impurity state
and the moir\'e potential gives rise to different patterns depending on the location. This interference
fully disappears when $\delta_J$ and $\delta_\Delta$ are switched off. 
We will show that our machine learning algorithm
will use the interference between the different locations of impurities to extract the values of the
Hamiltonian modulations.

We finally elaborate on the computational procedure to solve this model
in the limit of a single impurity in an otherwise pristine system.
For this purpose, we will use Green's function embedding
method.\cite{Lado2016,PhysRevResearch.2.033466,PhysRevMaterials.6.094010}.
The embedding method relies on extracting the Green's function
from the Dyson equation of the
defective system

\begin{equation}
    G_V(\omega) = [\omega - H_V - \Sigma(\omega)+i0^+]^{-1}
    \label{eq:gv}
\end{equation}

where $G_V(\omega)$ is the Green's function of the defective model, $H_V$ the Hamiltonian
of the defective unit cell, and $\Sigma(\omega)$ the selfenergy induced by
the pristine system.
The selfenergy $\Sigma(\omega)$ can
be obtained from the Dyson equation for the pristine model
$
    \Sigma(\omega) = \omega - H_0 - G_0(\omega)^{-1} +i0^+
    \label{eq:sigma}
$
with $H_0$ the Hamiltonian of the pristine unit cell,
where we take the Bloch's representation
of the pristine unit cell Green's function
$
    G_0(\omega) = 
    \frac{1}{(2\pi)^2} \int [\omega- H_{\mathbf{k}} + i0^+]^{-1}
    d^2\mathbf{k}
$
where $H_{\mathbf{k}}$ is the Bloch's Hamiltonian. 

The local density of states used as input for our
algorithm is obtained as

\begin{equation}
\rho (\mathbf r,\omega) = -\frac{1}{\pi} \sum_{s,\tau}
\langle \mathbf r,s,\tau | 
\text{Im} (G_V (\omega)) |
\mathbf r,s,\tau \rangle
\end{equation}

where $s$ and $\tau$ are the spin and Nambu indexes. 
The in-gap state $\epsilon_{imp}$ is located using an iterative algorithm
in the energy window of the topological superconducting gap.
From an experimental point of
view, a finite noise of the spectroscopy will be present, and thus the robustness of our algorithm will be important
with regards to its experimental implementation. We emulate the impact of noise in the real-space spectroscopy as 

\begin{equation}
\rho_{noisy} (\mathbf r) = \rho (\mathbf r) + \chi(\mathbf r)
\end{equation}
adding a noise background $\chi(\mathbf r) = \chi_0 \cdot \mathcal{R}(\mathbf r)$ to the local density of states (LDOS) $\rho (\mathbf r)$ where $\chi_0$ is the noise magnitude and $\mathcal{R}(\mathbf r) $ is a random uniform noise distribution defined in the interval $(-\langle \rho (\mathbf r) \rangle, \langle \rho (\mathbf r) \rangle)$ with $\langle \rho (\mathbf r) \rangle$ as mean value of the LDOS.

\subsection{Machine Learning}
\label{sec:ML}

\begin{figure}[t!]
\center
\includegraphics[width=.99\linewidth]{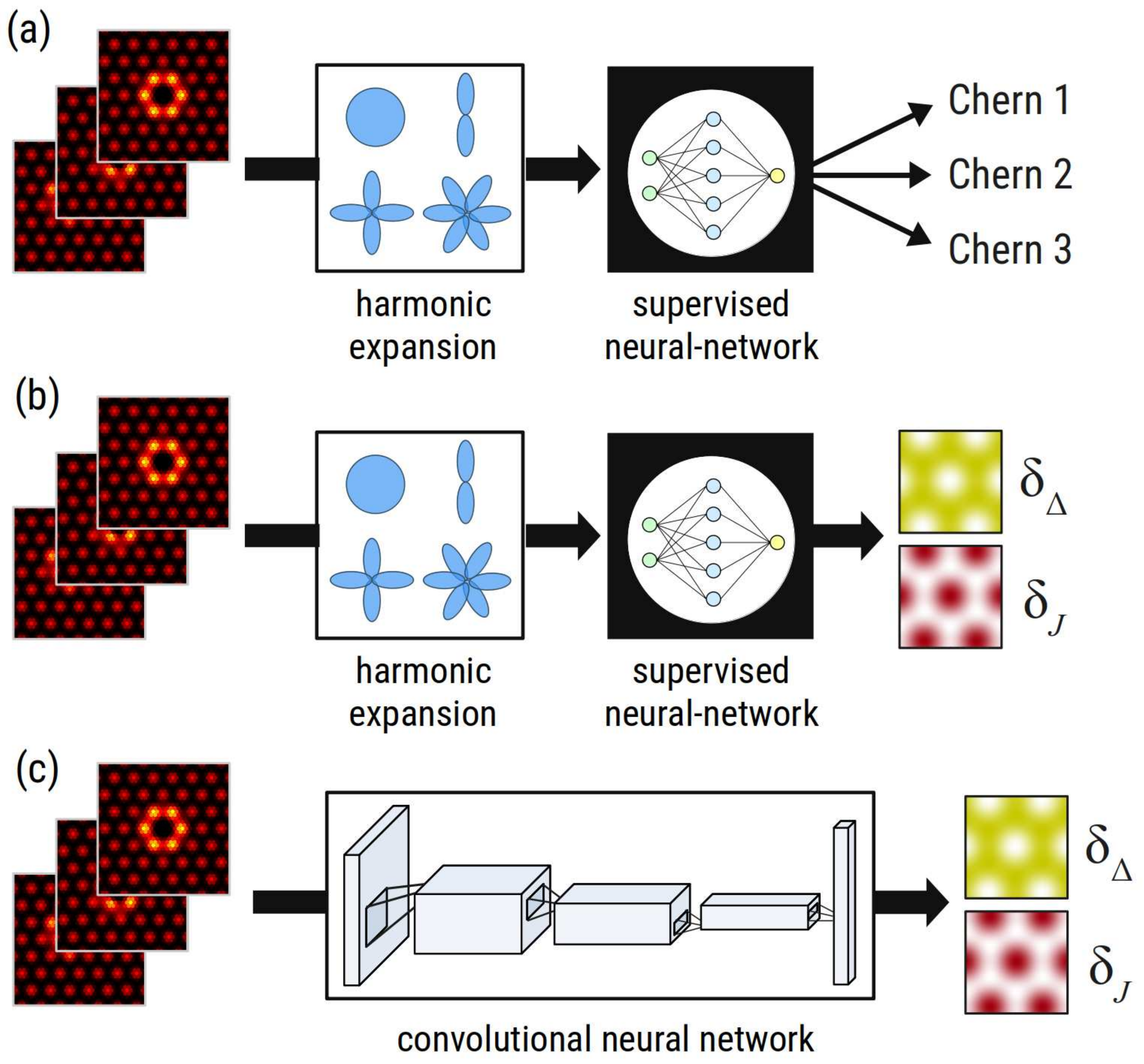}
\caption{ Schematic of the machine learning workflows. 
(a) Classification of the Chern number ($C=1,2,3$) with the radial expansion method and supervised NNs. 
(b) Hamiltonian learning (regression) of the exchange and superconducting modulation ($\delta_J$ and $\delta_\Delta$) with a feed-forward neural network and a radial expansion of the three LDOS.
(c) Hamiltonian learning of the exchange and superconducting modulation ($\delta_J$ and $\delta_\Delta$) with convolutional neural networks (CNNs). }
\label{fig:workflow}
\end{figure}

The main idea of our real-space impurity tomography is to extract the underlying Hamiltonian 
by using the LDOS for different impurity positions as inputs as shown in Fig.~\ref{fig:schematic}(f).
The extraction of the superconducting and exchange modulation is a non-trivial problem without a simple one-to-one correspondence to a single real-space spectroscopy.

We will use a physics-inspired approach to compress the information contained in the LDOS with a harmonic expansion (see Fig.~\ref{fig:workflow}(ab)). The coefficients of the power spectrum for each LDOS are then concatenated and fed into a fully-connected neural network (NN) as depicted in Fig.~\ref{fig:workflow}(b). The radial expansion works as follows. First, each individual LDOS is centered around the impurity position. Then, the harmonic expansion for the power spectrum is performed as

\begin{equation}
\centering
    c_{ln}=
    \sum_\alpha \rho (\mathbf r_\alpha) e^{i l \phi_\alpha} r_\alpha^n e^{-r_\alpha/\Lambda}       
\end{equation}

where $c_{ln}$ are the coefficients of the expansion, $\mathbf r_\alpha = r_\alpha(\cos \phi_\alpha,\sin \phi_\alpha,0)$ 
the atomic sites taking as the origin the impurity site, and $\Lambda$ is the typical localization length
of the state.\footnote{We take $\Lambda = 4a$, with $a$ the lattice constant in our calculations.}
In particular, we expand each LDOS into 24 complex $c_{ln}$ coefficients, i.e. the input dimension of the NN is 144 after concatenating the real and imaginary components of the coefficients for three LDOS.
We first train a fully-connected NN to perform classification of the Chern number of a given LDOS image which is compressed again into radial functions. The workflow is shown in Fig.~\ref{fig:workflow}(a). 
The NN architecture is shown in Table~\ref{tab:NN_class}. The loss function for this case is the \textit{categorical crossentropy loss} and a \textit{softmax-activation function} is used for the output layer.
The training consists of 100 epochs, with a batch size of 16. The optimization is performed with the Adam optimizer and a learning rate of $0.001$.
Afterward, we create a supervised architecture to predict the Hamiltonian parameter of the exchange and superconducting modulation ($\delta_{J}, \delta_{\Delta}$). For this regression task, the architecture of the NN is shown in Table~\ref{tab:NN_regression}.
We trained the NN for $100$ epochs with a batch size of 16. For the optimization of the weights, we are using the stochastic gradient descent algorithm and the Adam optimizer~\cite{adamOptim} with a learning rate of 0.001. The loss function is the mean squared error (MSE).
For the training and testing, we created LDOS for 2000 Hamiltonians for each Chern number where we varied the parameter $\delta_{J} \in [0, 2\lambda_R]$ and $\delta_{\Delta} \in [0, \lambda_R]$. 
This results in $6000$ samples which we divided into a training set of 5400 and a test set of 600 examples.
Finally, we take as a benchmark of our procedure a convolutional neural network (CNN)~\cite{lecun2015deep} architecture.
We take an analogous workflow as shown in Fig.~\ref{fig:workflow}(c), where
three LDOS with different impurity positions are fed into the CNN which predicts the modulation of the exchange or superconductivity as a regression task.
The architecture of the CNN is shown in Table~\ref{tab:CNN}, where three CNN networks are concatenated into a fully-connected NN and trained simultaneously with three input LDOS.  For the training, we used 50 epochs with a batch size of 16. 

It is finally worth noting that our machine learning approach allows us to exploit the spatial resolution of scanning tunneling microscopy. In the case of unconventional superconductors, the non-trivial interference between bound states provides information about the underlying electronic state, information that is often not accessible with other probes. Other approaches to Hamiltonian learning use the energy and entanglement spectra for Hamiltonian inference\cite{PhysRevB.97.075114}, or employing time-resolved measurements to extract Hamiltonian parameters \cite{PhysRevResearch.3.023246}.
However, performing those measurements for unconventional superconductors is greatly challenging,
whereas the measurements of the local density of states in our impurity tomography approach
are standard in scanning probe experiments in two-dimensional superconductors.

\section{Results and Discussion}

\begin{figure}[t!]
    \centering
    \includegraphics[width=\linewidth]{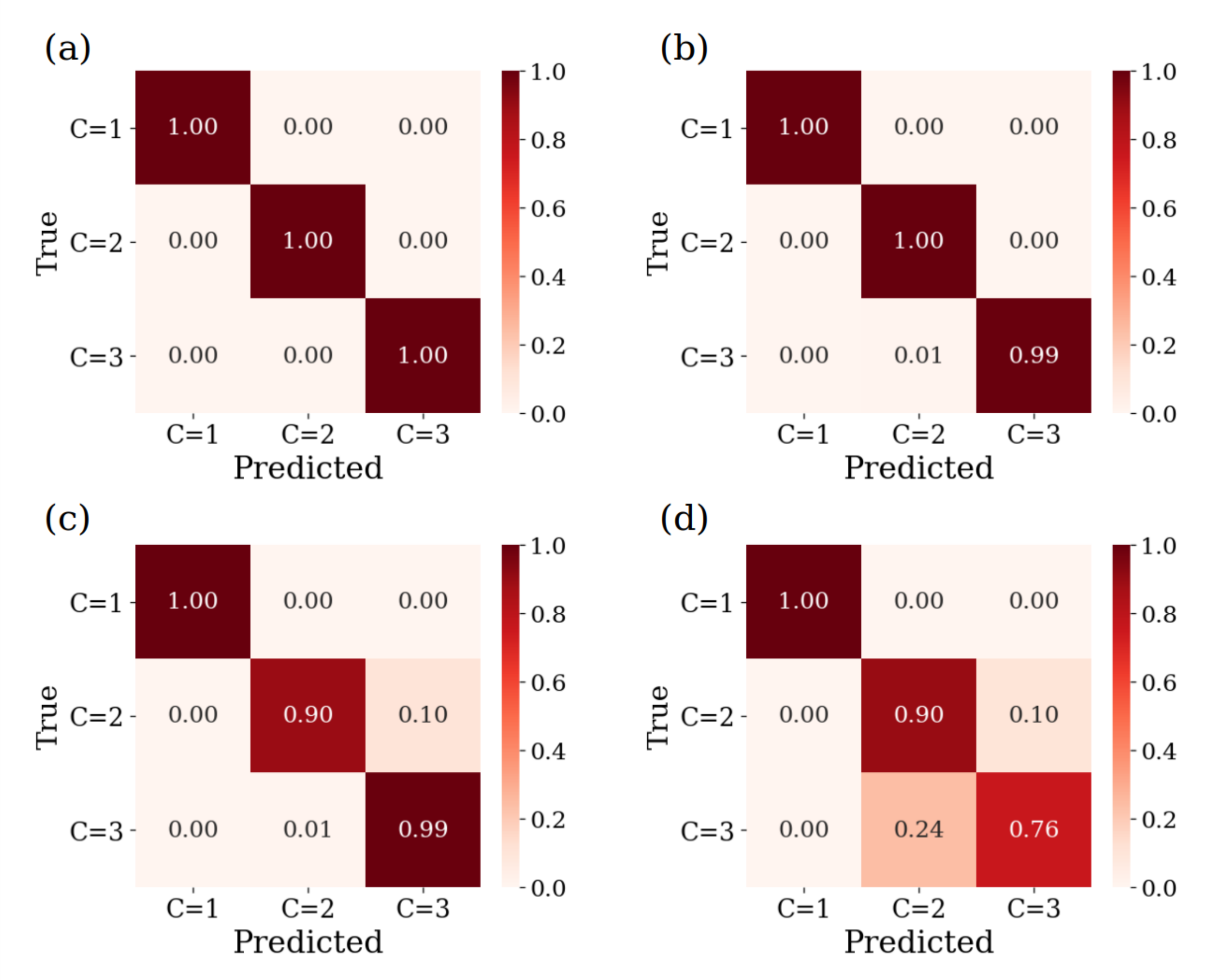}
    \caption{Extraction of the Chern number. Panels (a) and (b) show the prediction accuracy of C=1,2,3 when using three impurity positions as input for the NN for zero noise (a) and the maximum noise amplitude (b). Panels (c) and (d) show the prediction accuracy of $C=1,2,3$ when using one impurity position as input for the NN for zero noise (c) and the maximum noise amplitude (d).}
    \label{fig:confusion_matrix}
\end{figure}

\subsection{Chern number extraction}
The first problem we tackle is to determine the
Chern number from one or three LDOS with
different impurity locations.
The measurement of topological invariants in topological superconductors is a practical open challenge\cite{PhysRevB.97.115453,RodriguezNieva2019,PhysRevLett.124.226401,PhysRevB.102.054107,Kming2021,2019arXiv190103346C},
as due to their topological nature, local order parameters cannot be defined for these states.
The appearance of in-gap modes alone is not enough to assess the topological invariant 
of a superconductor, as in-gap states can appear in both trivial and topological superconductors\cite{RevModPhys.78.373}.
Interestingly, the inclusion of a moir\'e pattern leads to subtle changes in the spatial distribution
of in-gap states\cite{khosravian2022impurity}. As we will show below, the fine structure of the in-gap modes allows for extracting the
topological invariant of the underlying state.

\begin{figure}[t!]
    \centering
    \includegraphics[width=\linewidth]{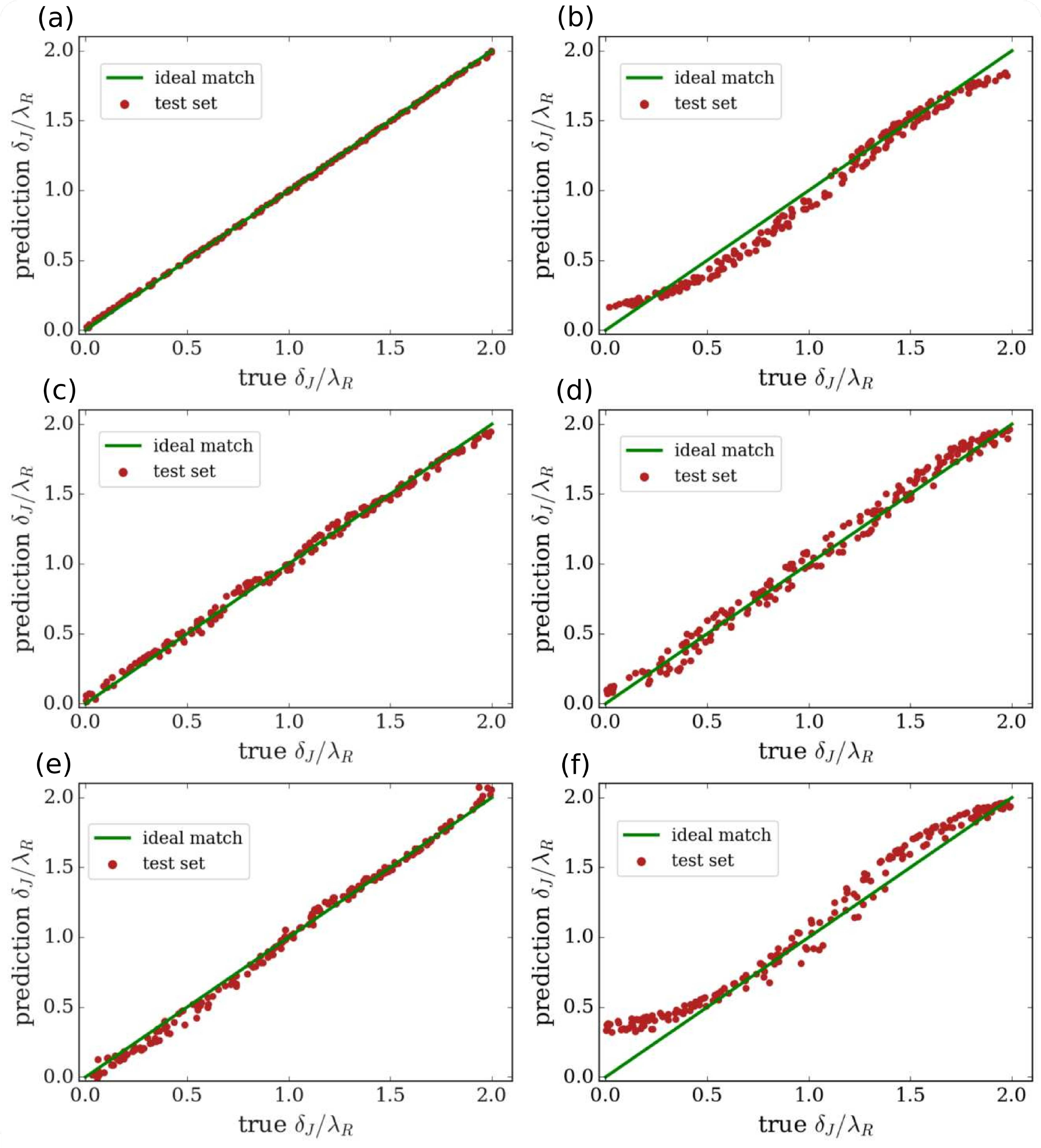}
    \caption{ Predictions of $\delta_{J}$ with a harmonic expansion fully-connected architecture
    (a,c,e) compared to the convolutional architecture predictions (b,d,f) for Chern numbers $C=1$ (a,b), $C=2$ (c,d), and $C=3$ (e,f).
    It is observed that the harmonic expansion fully-connected architecture provides slightly more accurate predictions than the
    convolutional architecture.}
    \label{fig:exchange_results}
\end{figure}

In the following, we elaborate on the procedure to extract the topological invariant solely from
the real-space spectroscopy images of single impurities. The input of the algorithm consists of the real-space spectroscopy
for three impurity locations, and its output is the topological invariant.
To extract the topological invariant, we trained a NN to perform the classification into the 3 non-trivial Chern numbers. The results are shown in Fig.~\ref{fig:confusion_matrix} in the form of confusion matrices for three Fig.~\ref{fig:confusion_matrix}(a,b) and one LDOS Fig.~\ref{fig:confusion_matrix}(c,d) as input. For one LDOS as input, we achieve an average accuracy of 96.3$\%$ for the test data with zero noise (c) and 88.7$\%$ with the maximum noise amplitude of $\chi_0=0.4$. 
The confusion matrix shows that $C=1$ can be predicted with 100$\%$ accuracy in both cases, the NN is only confused in the prediction between $C=2$ and $C=3$.
Taking three real-space spectroscopies with different impurity positions as input, we obtain a testing accuracy of 100$\%$ for zero noise (a) and  99.7$\%$ for the maximum noise amplitude.
This highlights that combining three real-space spectroscopies for different impurity locations substantially increases the accuracy of the topological invariant predictions.
These results show, that it is possible to extract the Chern number by just one LDOS with a random
impurity position even in the case of very noisy data.

\begin{figure}[t!]
    \centering
    \includegraphics[width=\linewidth]{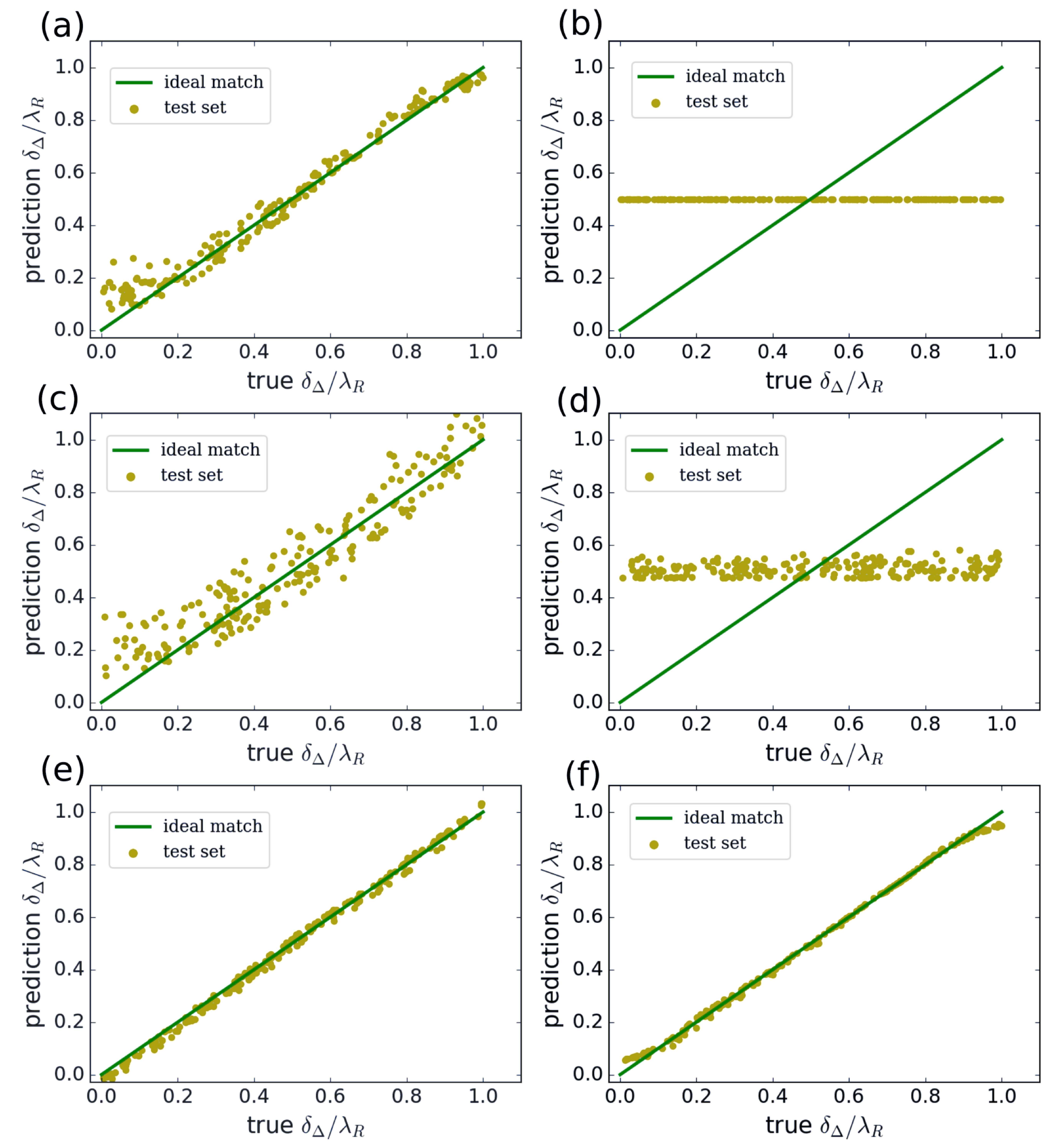}
    \caption{ Predictions of $\delta_{\Delta}$ with a harmonic fully-connected architecture (a,c,e) compared to a convolutional architecture (b,d,f) for Chern numbers $C=1$ (a,b), $C=2$ (c,d), and $C=3$ (e,f).
    It is observed that the harmonic fully-connected architecture provides more accurate predictions than the
    convolutional architecture, in particular for $C=1,2$ where the convolutional architecture fails.}
   \label{fig:sc_results}
\end{figure}

\subsection{Exchange field extraction}
In the following, we show how the real-space impurity tomography allows extracting the values of the moir\'e Hamiltonian parameters.
We start with the modulation of the exchange coupling, which is intuitively the parameter that will impact the
real-space spectroscopy in the strongest way. This stems from the fact that the local exchange coupling creates in-gap
Yu-Shiba-Rusinov modes inside the original superconducting gap $\Delta$\cite{PhysRevLett.115.197204,Kim2018,Feldman2016,PhysRevLett.121.196803,Schneider2021,PhysRevB.100.214504,Kezilebieke2020,PhysRevB.102.104501,Kezilebieke2022}, 
and these modes form the low energy electronic
structure leading to a topological superconducting state. As a result, low energy bands reflect the periodicity
of the exchange modulation, and their associated in-gap modes in the presence of
impurities inherit the same dependence\cite{Kezilebieke2022,PhysRevMaterials.6.094010}.

We take as input the real-space spectroscopy for three
impurity locations, leading as output the modulation strength of the exchange coupling in the system.
We train different architectures for each Chern number, including convolutional
neural networks of the full spectroscopy and the harmonic expansion of the impurity states.
The results, shown in Fig.~\ref{fig:exchange_results} demonstrate that real-space spectroscopy
of the impurities allows extracting the modulation of the exchange coupling with both procedures.
Specifically, it is shown that both convolutional and harmonic expansion obtain a substantial accuracy in the extraction, specifically leading to a typical error of $\mathcal{E} [\delta_J ] = 0.01$ 
for the convolutional algorithm and 
$\mathcal{E} [\delta_J] =0.0072 $ for the harmonic expansion. We observe that the harmonic algorithm
becomes slightly more accurate than the convolutional method.
Interestingly, the difference between both methods becomes much more dramatic
in the extraction of superconducting order values and modulations.

\begin{figure}[t!]
    \centering
    \includegraphics[width=.99\linewidth]{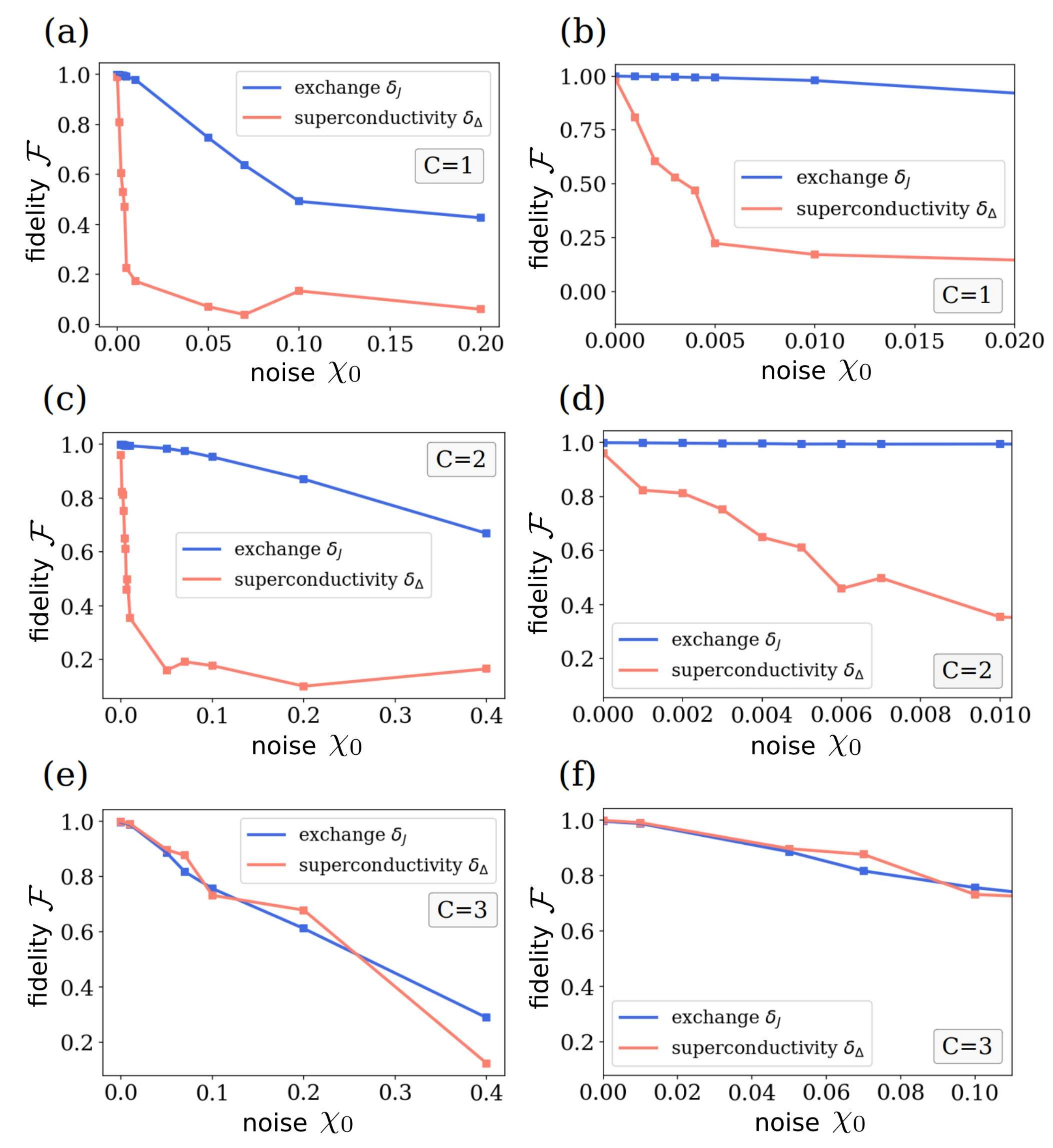}
    \caption{Analysis of the correlation between the predicted and real values of the exchange and superconducting modulation
    as a function of the noise level for $C=1$ (a,b), $C=2$ (c,d), and $C=3$ (e,f).
    Panels (a,c,e) correspond to a larger noise range, whereas (b,d,f) correspond to the low noise limit. 
    It is observed that exchange modulations $\delta_J$ are more resilient to noise than superconducting
    modulation predictions $\delta_\Delta$.
    }
    \label{fig:noise}
\end{figure}

\subsection{Superconducting order extraction}
We now move to consider the extraction of the local superconducting
order $\Delta (\mathbf r)$ with real-space impurity tomography. It is interesting to note that the impact of
modulation of $\Delta$ in the low energy states is expected to be more subtle
than the exchange. While the energy location of a Yu-Shiba-Rusinov strongly
depends on $J$, the value of the superconducting gap only leads to proportional
renormalization of the energy\cite{RevModPhys.78.373}.

We show in Fig.~\ref{fig:sc_results} the extraction of the superconducting modulation using the harmonic expansion with supervised learning (Fig.~\ref{fig:sc_results}(ace))
and the convolutional neural network architecture (Fig.~\ref{fig:sc_results}(bdf)),
for $C=1$ (Fig.~\ref{fig:sc_results}(ab)), $C=2$ (Fig.~\ref{fig:sc_results}(cd))
and $C=3$ (Fig.~\ref{fig:sc_results}(ef)). In the case $C=3$ (Fig.~\ref{fig:sc_results}(ef)),
we observe that both architectures are capable of predicting the superconducting modulation.
Interestingly, for $C=1$ (Fig.~\ref{fig:sc_results}ab) and $C=2$ (Fig.~\ref{fig:sc_results}cd),
the convolutional architecture is incapable of predicting correct superconducting modulations,
whereas the fully-connected harmonic expansion successfully predicts it. 
Typical errors of the NNs are $\mathcal{E} [\delta_\Delta] =0.30, 0.29, 0.008$ for $C=1,2,3$ for
CNN and $\mathcal{E} [\delta_\Delta] =0.03, 0.024, 0.008$  for $C=1,2,3$ for harmonic expansion with the fully-connected NN.
This phenomenology should be contrasted with the exchange modulation, where both architectures showed similar performance. 
The success of the fully-connected harmonic expansion and failure the CNN architecture demonstrates that the extraction of superconducting modulations represents a much more challenging problem than the exchange modulation. 
In particular, our harmonic expansion shows that such a physically-motivated procedure allows extracting Hamiltonian parameters stemming from highly subtle changes in the real-space spectroscopy.

\subsection{Noisy Hamiltonian extraction}

\begin{figure}[t!]
    \centering
    \includegraphics[width=1.0\linewidth]{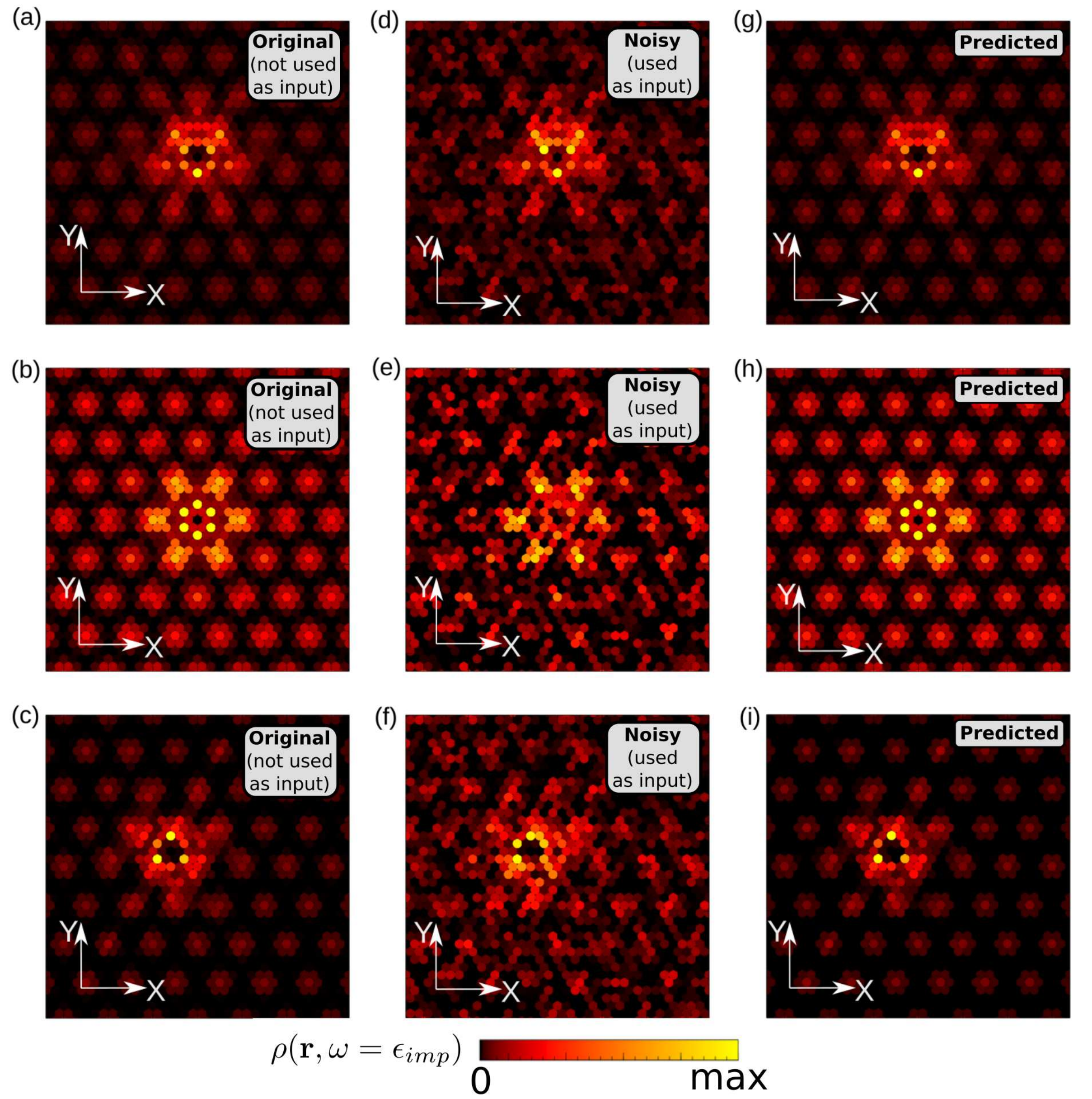}
    \caption{Local density of states of the in-gap state for a single impurity in a moir\'e superlattice with $C=3$ for three different locations for (a)-(c). Panels (d)-(f) show the real-space spectroscopy with $\chi_0=0.2$ noise amplitude, which is used as input
    for the machine learning algorithm. Panels
    (g)-(i) show the real-space spectroscopy associated to the predicted Hamiltonian by the neural network.
    }
    \label{fig:noise_ldos}
\end{figure}

In the following, we address the extraction of the Hamiltonian in the presence of noisy data. 
For the sake of concreteness, we focus on the harmonic fully-connected architecture, as 
the superconducting modulation can be predicted with that one. 
We show in Fig.~\ref{fig:noise} the evolution of the correlation between predictions and true values
as a function of increasing noise. We define prediction fidelity as

\begin{equation}
\begin{split}
    & \mathcal{F} (\delta^{pred}, \delta^{true}) = \\[1mm] 
    & \frac{
    |\langle \delta^{pred} \delta^{true} \rangle - \langle \delta^{true} \rangle \langle \delta^{pred} \rangle|
    }{\sqrt{(\langle (\delta^{true})^2 \rangle - \langle \delta^{true} \rangle^2) (\langle (\delta^{pred})^2 \rangle - \langle \delta^{pred} \rangle^2) }}
\end{split}
\end{equation}
is defined in the interval $\mathcal{F} \in [0,1]$, where $\mathcal{F}=1$ corresponds
to the best prediction accuracy $\delta^{pred} = \delta^{true}$
and $\mathcal{F}=0$ corresponds to no predictive accuracy.
Focusing first on the exchange modulation shown in Fig.~\ref{fig:noise}, we observe that even substantial noise allows
to infer the value of its modulation for the Chern numbers. The robustness of the exchange modulation extraction 
is consistent with the fact that the neural network architecture was capable of providing faithful predictions of the parameter.
This resilience to noise stems from the strong dependence of the Yu-Shiba-Rusinov states on the local exchange, which
directly impacts the real-space profile of the spectroscopy. The prediction of the superconducting modulation represents
however a bigger challenge as shown in Fig.~\ref{fig:noise}. In the cases $C=1,2$ (Fig.~\ref{fig:noise}(abcd), the most challenging
ones, we observe that magnitudes of noise around $\chi_0 = 0.005$ break down the predictions. In stark contrast, systems with $C=3$ are resilient to levels of noise magnitudes up to $\chi_0 = 0.2$ as shown in Fig.~\ref{fig:noise}(ef). The difference in the robustness 
between $C=1,2$ and $C=3$ can be rationalized by recalling that the convolutional neural network architecture is not capable
of providing predictions for $C=1,2$, but it gave reliable predictions for $C=3$.
Interestingly, 
despite the superconducting modulation creating
a very subtle impact in the real-space spectroscopy, our algorithm is capable of extracting the modulation
amplitude from noisy data for different Chern numbers.
Finally, we show in Fig.~\ref{fig:noise_ldos} a comparison between the spectroscopies for
the pristine data (Fig.~\ref{fig:noise_ldos}abc), the noisy data provided as input to our algorithm
(Fig.~\ref{fig:noise_ldos}def) and the spectroscopy computed with the parameters predicted by the Hamiltonian extraction
(Fig.~\ref{fig:noise_ldos}ghi). It is observed that the spectroscopy predictions are nearly indistinguishable from
the original ones, showing that our algorithm is capable of solving the inverse problem of inferring the Hamiltonian
from the spectroscopy.

\section{Conclusions}

To summarize, we have shown that interference effects of in-gap states allow extracting Hamiltonian moir\'e parameters
and topological invariants in an artificial topological superconductor from real-space spectroscopy patterns. 
Our approach is based on a supervised learning procedure that
exploits the patterns obtained in in-gap states for different impurity locations simultaneously.
We showed that our technique
can be readily implemented with a convolution neural network architecture that combines several impurity locations
simultaneously. Furthermore, we showed that by leveraging an orbital expansion of the impurity modes,
a more robust machine-learning approach can be developed. Specifically, our procedure using a harmonic expansion
is capable of extracting both parameter values and their modulations even in the presence of noise.
We demonstrate that the combination of different locations dramatically increases the accuracy of Hamiltonian
parameter inference, exemplifying how parameter inference benefits from
features that our algorithm extracts from several impurity locations.
Our results establish a machine learning methodology that exploits local impurity engineering to extract
non-trivial parameters in artificial topological systems. Our demonstration exploits data directly accessible
in scanning tunneling microscopy experiments and is robust in the presence of noise, establishing
a realistic method to extract Hamiltonian parameters from readily accessible experimental data in complex quantum materials.

\textbf{Acknowledgements}:
We acknowledge the computational resources provided by
the Aalto Science-IT project,
and the financial support from the
Academy of Finland Projects Nos. 331342, 336243 and 358088,
the Jane and Aatos Erkko Foundation.
We thank P. Liljeroth, S. Kezilebieke and
T. Ojanen for useful discussions.

\appendix

\section{Appendix: Network Architectures}
\label{App:NN}

\begin{table}[h]
    \centering
    \caption{Architecture of the fully-connected NN of the expansion method. Used for the classification of the Chern number.}
    \label{tab:NN_class}
        \begin{tabularx}{.48\textwidth}{ |X X| }
          \hline
          \multicolumn{2}{|c|}{fully-connected NN (classification)} \\
          \hline
          \hline
          Layer & Output shape \\
          \hline 
          \textbf{harmonic expansion} & \hspace{16mm} \textit{preprocessing} \\
          InputLayer & (48 $\cdot$ 3)  \hspace{2mm} \textit{48 params / LDOS}\\
          Dense & (200) \\
          Dropout & (200) \\
          Dense & (100) \\
          Dropout & (100) \\
          Dense & (20) \\
          Dropout & (20) \\
          Dense & (3) \\
          \hline
          total parameters & 51,183 \\
          \hline
        \end{tabularx}
\end{table}

\begin{table}[h]
    \centering
    \caption{Architecture of the fully-connected NN of the expansion method. Used for the regression of the Hamiltonian parameters.}
    \label{tab:NN_regression}
        \begin{tabularx}{.48\textwidth}{ |X X| }
          \hline
          \multicolumn{2}{|c|}{fully-connected NN (regression)} \\
          \hline
          \hline
          Layer & Output shape \\
          \hline 
          \textbf{harmonic expansion} & \hspace{16mm} \textit{preprocessing} \\
          InputLayer & (48 $\cdot$ 3)  \hspace{2mm} \textit{48 params / LDOS}\\
          Dense & (300) \\
          Dropout & (300) \\
          Dense & (2) \\
          \hline
          total parameters & 15,302 \\
          \hline
        \end{tabularx}
\end{table}

\begin{table}[h]
    \centering
    \caption{Architecture of the CNN. Used for the regression of the Hamiltonian parameters.}
    \label{tab:CNN}
        \begin{tabularx}{.48\textwidth}{ |X X| }
          \hline
          \multicolumn{2}{|c|}{CNN} \\
          \hline
          \hline
          Layer & Output shape \\
          \hline 
          \textbf{combine three CNNs:} & \\
          InputLayer & (55,55,1) \\
          Conv2D & (55, 55, 32) \\
          MaxPooling2D & (27, 27, 32) \\
          Conv2D & (27, 27, 64) \\
          MaxPooling2D & (13, 13, 64) \\
          Flatten & (10816) \\[0.5mm]
          \textbf{fully-connected NN} & \\
          Concatenate & (32448) \hspace{2mm} \textit{combine 3 CNNs} \\
          Dense & (200) \\
          Dense & (50) \\
          Dense & (1) or (2) \\
          \hline
          total parameters & 6,556,349 \\
          \hline
        \end{tabularx}
\end{table}

\bibliography{references}

\end{document}